\documentclass[grlga]{aguplus}                     

 \usepackage{graphicx}

\printfigures              
\afour                     

\lefthead{Hansen et al}
\righthead{Stress of Rough Surfaces}
\received{March xx, 2000}
\journalid{GRL}{March 2000}
\articleid{1}{4}
\paperid{xxxxxxxx}
\ccc{0000-0000/00/xxxxxxxx\$05.00}
\cpright{AGU}{2000}
\authoraddr{G.\ George Batrouni, Institut Non-Lin\'eaire de Nice, 
UMR CNRS 6618, Universit{\'e} de Nice--Sophia Antipolis, 1361 Route
des Lucioles, F--06560 Valbonne, France}
\authoraddr{Alex Hansen, Department of Physics, Norwegian University of Science
and Technology, N--7491 Trondheim, Norway}
\authoraddr{Fernando  A.\ de Oliveira, ICCMP, Universidade de Bras{\'\i}lia, 
CP 04513, 70919--970 Bras{\'\i}lia--DF, Brazil}
\authoraddr{Jean Schmittbuhl, Laboratoire de G{\'e}ologie, UMR CNRS 8538,
Ecole Normale Sup{\'e}rieure, 24, rue Lhomond, F--75231 Paris C{\'e}dex 05, 
France}

\slugcomment{Submitted to Geophysical Research Letters}
\begin{document}
\title{Normal Stress Distribution of Rough Surfaces in Contact} 

\author{Alex Hansen,\altaffilmark{1,}\altaffilmark{2}
Jean Schmittbuhl,\altaffilmark{3} 
G.\ George Batrouni,\altaffilmark{4} and Fernando A.\ de Oliveira}
\affil{International Center for Condensed Matter Physics, 
Universidade de Bras{\'\i}lia, Brazil} 

\altaffiltext{1}{Present address: NORDITA, Blegdamsvej 17, DK-2100 
Copenhagen, Denmark}

\altaffiltext{2}{Permanent address: Department of Physics, Norwegian 
University of Science and Technology, N-7491 Trondheim, Norway}

\altaffiltext{3}{Permanent address: Laboratoire de G{\'e}ologie, UMR CNRS 8538,
Ecole Normale Sup{\'e}rieure, 24, rue Lhomond, F--75231 Paris C{\'e}dex 05, France}

\altaffiltext{4}{Permanent address: Institut Non-Lin\'eaire de Nice, 
UMR CNRS 6618, Universit{\'e} de Nice--Sophia Antipolis, 1361 Route
des Lucioles, F--06560 Valbonne, France}
\begin{abstract}
We study numerically the stress distribution on the interface between
two thick elastic media bounded by interfaces that include spatially
correlated asperities. The interface roughness is described using the
self-affine topography that is observed over a very wide range of
scales from fractures to faults.  We analyse the correlation
properties of the normal stress distribution when the rough surfaces have
been brought into full contact. The self affinity of the rough
surfaces is described by a Hurst exponent $H$.  We find that the
normal stress field is also self affine, but with a Hurst exponent
$H-1$. Fluctations of the normal stress are shown to be important,
especially at local scales with anti-persistent correlations.
\end{abstract}
\begin{article}
  
Theories describing the elastic properties of two media in contact
through rough surfaces have important applications in a wide range of
geophysical problems, such as earthquakes, fracture, fluid permeability
or rock friction \citep{Sc90}. Asperities exist at all
scales: grain roughness is relevant for closure of rock joints
\citep{Br-Sc86} and seamounts might induce large scale stress
fluctuations along subduction slabs \citep{Dm-Zh-Ri}. Whatever the scale
of the asperities in contact is, when they are attached to an elastic
medium and are loaded, they interact and concentrate high stresses.
Friction properties of an interface are very dependent on the
heterogeneities of the normal stresses \citep{Di-Ki}. At fault scale,
residual stresses resulting from asperity squeeze might be responsible
for heterogeneities of the dynamic stress field and influence the
earthquake propagation \citep{Bo-Ca-Co}.
  
\begin{figure}
  \figurewidth{35pc}
  \figbox*{}{}{\includegraphics[width=35pc,angle=268]{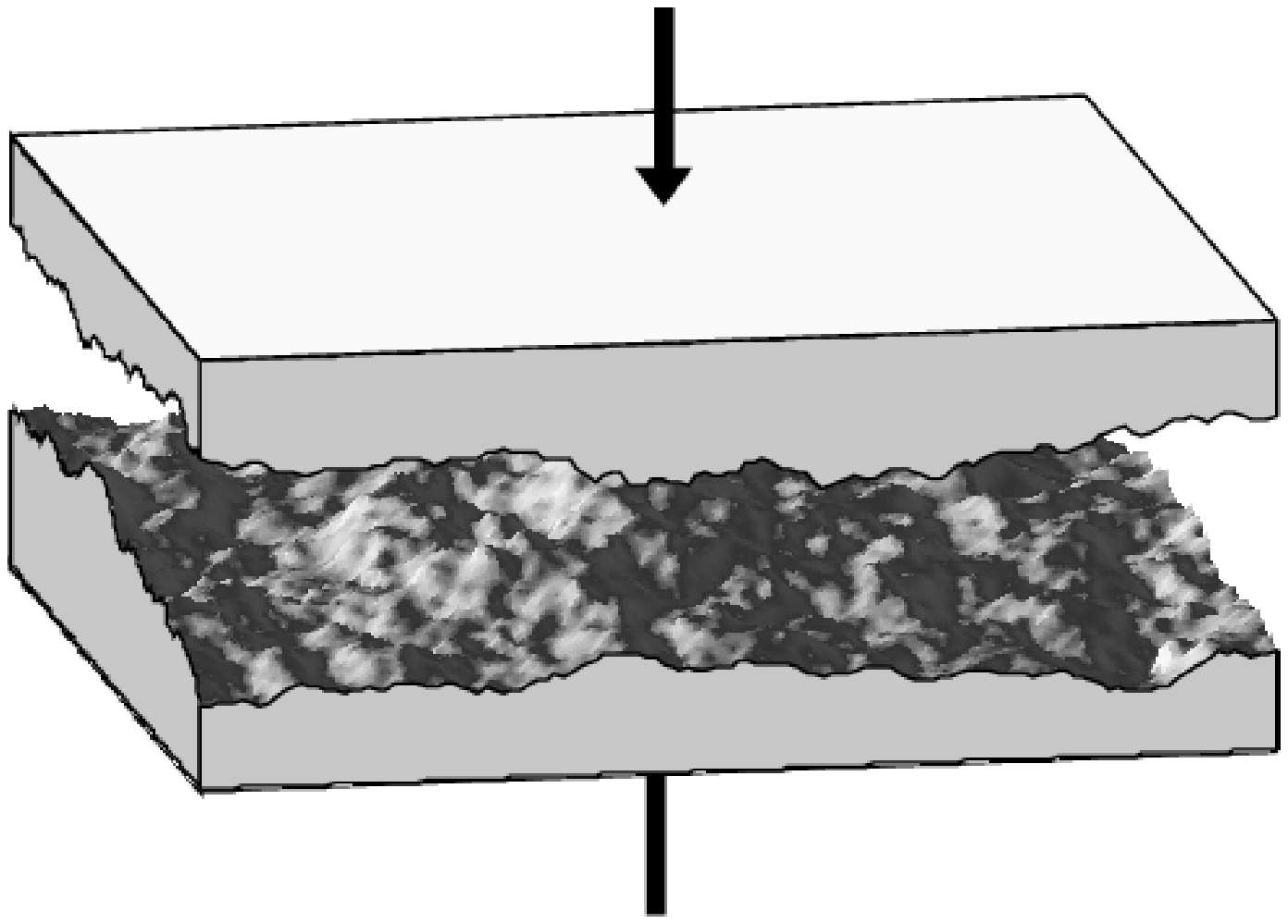}}
  \caption[]{\label{fig:realsurf} A rough granite surface produced by
  cleavage.  The size is $512\times 512$ points. Physical dimensions are
  $12.8 {\rm mm}\times 12.8{\rm mm}$.
    }
\end{figure}

We consider in this letter the normal component of the stress field
$\sigma_N$ that appears on the interface between two elastic media with
rough surfaces (see Figure~\ref{fig:realsurf}) \citep{Sa96}. We assume
that possible local plastic deformations where the elastic yield
stress of asperities is overcome, have a negligible effect on the stress
distribution along the interface. The roughness is assumed to be {\it
self affine\/}. The surface is described as $h(x,y)$.  Rescaling the
two coordinates $x\to\lambda x$ and $y\to \lambda y$, necessitates a
rescaling of the height $h\to \lambda^H$ \citep{Fe88}.  The surface is
then self affine with a Hurst exponent $H$.  In Figure
\ref{fig:realsurf} an example of surfaces with this property is shown.

This surface was obtained from an impact fracturing of a 
granite block
(25cm$\times$25cm) \citep{Lo-Sc}.  A wide range of experimental data
\citep{Br-Sc85a,Po-Tu-Br-Bo-Sc,Bo90,Ma92,Sc-Sc-Sc} support 
the hypothesis that not only are surfaces produced by brittle fracture
self affine, but their Hurst exponent generally equals $H=0.80$
independently of the material \citep{Bo97}.  We have measured the
roughness of the granite block using a profilometer and analysed the
spatial correlations of the surface with the average wavelet
coefficient technique \citep{Me97,Si-Ha-Ne}.  This consists in wavelet
transforming each one-dimensional trace $h(x,y={\rm const})$ using the
Daubechie-12 wavelet basis and averaging over the wavelet coefficients
at each length scale $b=2^k$, where $k$ is an integer.  If the trace
is self affine, the averaged wavelet coefficients scale as
\begin{equation}
A_b \sim b^{H+1/2}\;,
\end{equation}
where $A_b$ is an average over the position of the
wavelet.

\begin{figure*}
  \figurewidth{35pc}
  \figbox*{}{}{\includegraphics[width=25pc,angle=270]{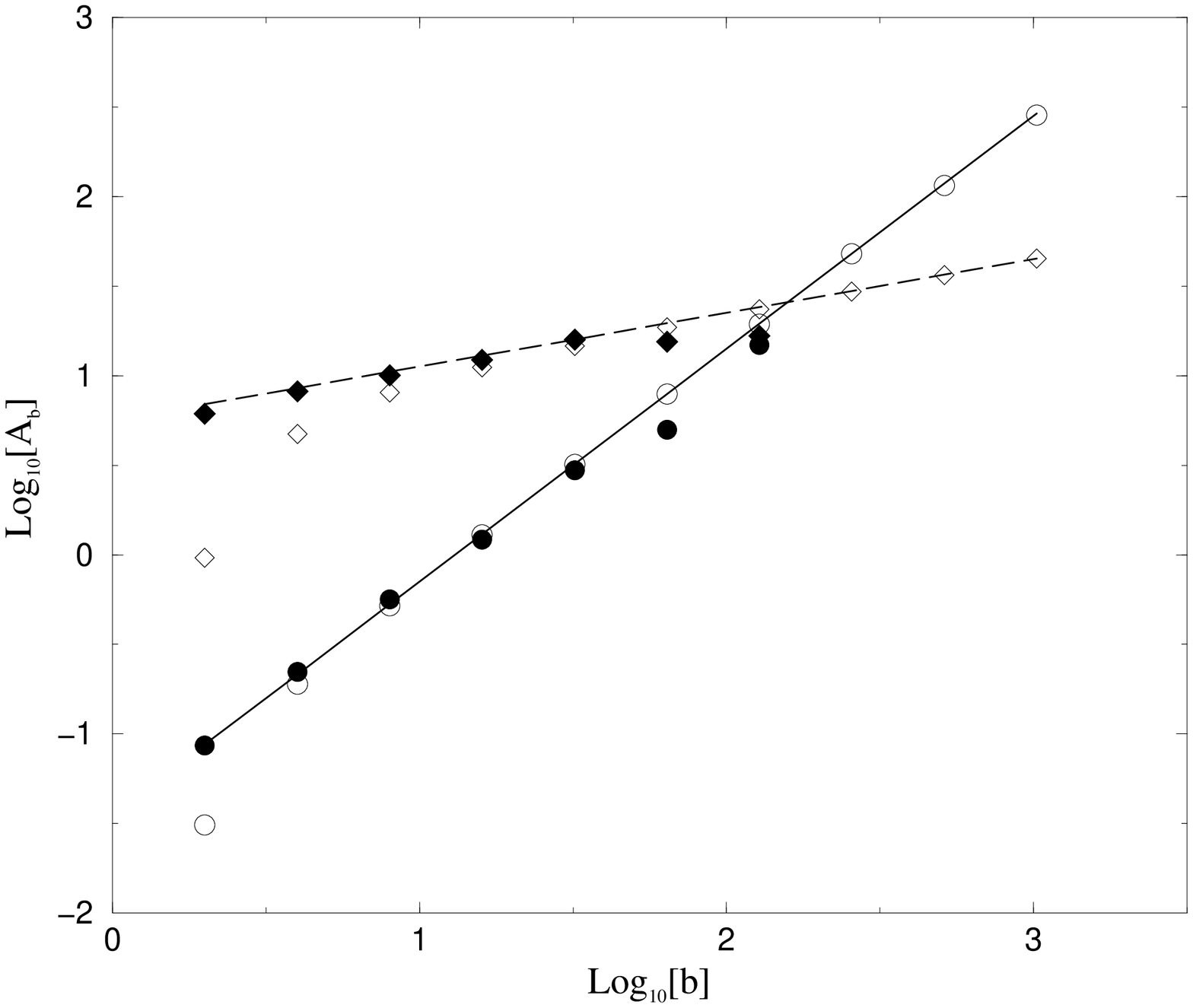}}
  \caption[]{\label{fig:wavesurf} The average wavelet coefficient $A_b$ 
    for the fracture surface of Figure \ref{fig:realsurf} as a function of
    length scale $b$ is shown as filled circles. Empty circles correpond to
    the average over 100 synthetic surfaces of 2048$\times$2048. The slope
    of the straight solid line is $1.30$, corresponding to a Hurst exponent
    $H=0.80$.  Average wavelet coefficient $A_b$ for the stress field shown
    in Figure \ref{fig:realsurfstress} as a function of length scale $b$
    using filled diamonds. Empty diamonds describe the analysis of the
    squeeze of the synthetic surfaces.  The slope of the dashed straight
    line is $0.30$, corresponding to a Hurst exponent $H_\sigma=-0.2$.  }
\end{figure*}

We show in Figure \ref{fig:wavesurf} the average wavelet coefficients {\it
  vs.\/} $b$ for the granite surface of Figure \ref{fig:realsurf}.  The
slope of the least-squares fit is $1.30\pm0.02$, giving $H=0.80\pm0.02$.

Integrating the Lam{\'e} equations for an infinite block limited by an
infinite plane $(x,y,z=0)$, gives the Green function $G$ for the
deformational response $u$ in the $z$ (vertical) direction at a point
$(x_u,y_u)$ in the plane $z=0$ from a distribution $f(x_f,y_f)$ of applied
forces in
the vertical direction:
\begin{equation}
u(x_u,y_u)=\int\int{G(x_u-x_f,y_u-y_f)f(x_f,y_f)}dx_f dy_f\;,
\label{eq:green1}
\end{equation}
where \citep{La-Li58}
\begin{equation}
G(x_u-x_f,y_u-y_f)=\frac{1-s^2}{\pi E}\ \frac{1}{r}\;.
\label{eq:green2}
\end{equation}
Here $r=\sqrt{(x_u-x_f)^2+(y_u-y_f)^2}$, $E$ is the elastic constant
and $s$ is the Poisson ratio.  Deformation also occurs within the
$(x,y)$-plane when vertical force is applied.  However, these fall of
as $1/r^2$.  Consequently, we ignore them compared to the deformation
in the vertical direction.

We are interested in rough self-affine surfaces. However, with a Hurst
exponent $H<1$, the surfaces are asymptotically flat.  This can be
easily seen by calculating the rms fluctuations of the surface,
$w^2=\langle (h-\langle h\rangle)^2\rangle_{L\times L}$, where
$\langle\cdots\rangle_{L\times L}$ is an average over an area of size
$L\times L$ of the plane $(x,y)$.  When the surface is self affine, we
have $w\sim L^H$ leading to $w/L\sim 1/L^{1-H} \to 0$ as
$L\to\infty$. Thus, it is asymptotically flat.  As we are primarily
interested in the scaling properties of the normal component of the
stress field $\sigma_N$ on large scales, it is a good approximation to
use the flat-surface Green function, (\ref{eq:green2}).  This
approximation also leads to the force component in the vertical
direction being directly proportional to the normal stress.

The problem we have set out to study is that of two self-affine rough
surfaces in full contact.  However, assuming that one of the surfaces
is rough and infinitely hard, and the other elastic and initially
flat, we find the same normal stress field at the interface as in the
original problem within the approximation using the flat-surface Green
function, (\ref{eq:green2}) and using the {\it composite} topography
introduced by \cite{br-Sc85b} ({\it i.e.} the sum of both
topographies).  We will, therefore, study this second problem since it
is easier to implement numerically.

When the two surfaces are in full contact, the deformation field $u$
will be equal to minus the local height, $u=-h$, when in-plane
deformations are ignored.  Thus, the deformation field is self-affine,
with a Hurst exponent $H$.

Since Eq.\ (\ref{eq:green1}) is linear, simple scaling arguments tell
us how the force field $f$ scales given that $u$ is self
affine.\footnote{If the underlying equations were not linear, much
more powerful methods to determine the scaling behavior would be
necessary such as functional renormalization
\citep{Ba-St}.}  If we scale $(x_u,y_u)\to (\lambda x_u,\lambda y_u)$ and
$(x_f,y_f)\to (\lambda x_f,\lambda y_f)$, Eqs.\ (\ref{eq:green1}) and
(\ref{eq:green2}) immediately gives the scaling relations
\begin{equation}
\left\lbrace
\begin{array}{l}
\Delta x \to \lambda \Delta x\;,\\
\Delta y \to \lambda \Delta y\;,\\
u\to \lambda^H u\;,\\
G\to \lambda^{-1} G\;,\\
f\to \lambda^{H-1} f\;,
\end{array}
\right.
\label{eq:xyuf}
\end{equation}
with $\Delta x=x_u-x_f$ and $\Delta y=y_u-y_f$.  Thus, the force field
$f$ is self affine with a Hurst exponent equal to $H_\sigma = H-1$.

\begin{figure}
  \figurewidth{35pc}
  \figbox*{}{}{\includegraphics[width=35pc,angle=0]{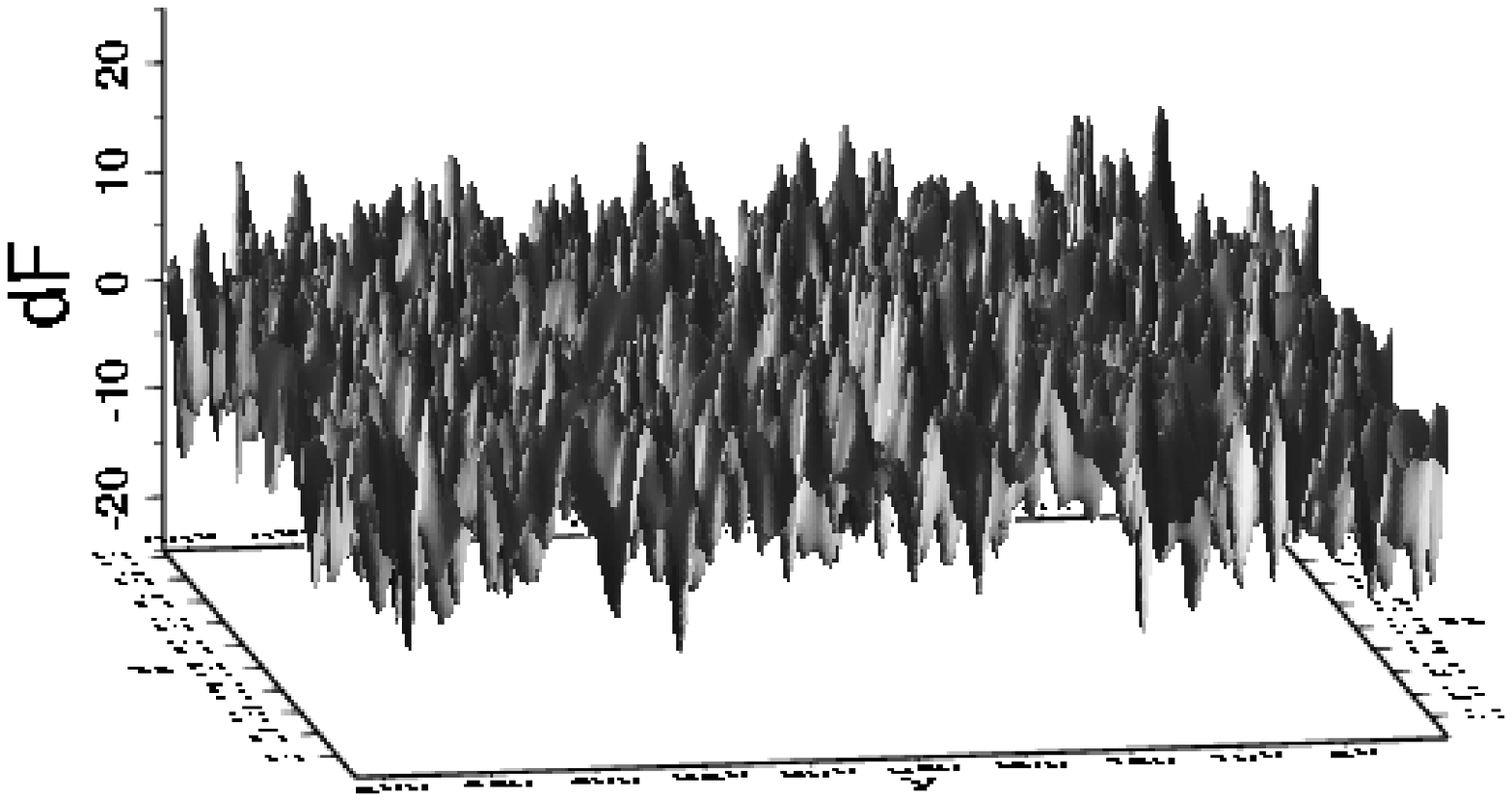}}
  \caption[]{\label{fig:realsurfstress} The numerically calculated
  stress field on the interface
  between the granite block of Figure \ref{fig:realsurf} when brought in
  complete contact with an infinitely hard flat surface.
    }
\end{figure}

In order to demonstrate the validity of Eq.\ (\ref{eq:xyuf}), we solve
Eq.\ (\ref{eq:green1}) numerically for $f$.  This is done in Fourier
space (using FFTs \citep{St-Ka}) since the Green function is diagonal
there.  We start out by defining the Green function on the $L\times L$
square lattice as follows. For each node $(i,j)$, we define
$r_1=((i-1)^2 +(j-1)^2)^{1/2}$, $r_2=((i-1)^2+(L+1-j)^2)^{1/2}$,
$r_3=((L+1-i)^2+(j-1)^2)^{1/2}$, and
$r_4=((L+1-i)^2+(L+1-j)^2)^{1/2}$.  The Green function on this lattice
is then given by
\begin{equation}
G_{(i,j)}=
\frac{1-s^2}{4\pi E} 
\left[\frac{1}{\max(r_1,\epsilon)}+\frac{1}{r_2}+
\frac{1}{r_3}+\frac{1}{r_4}\right]\;.
\label{eq:green3}
\end{equation}
Thus, the singularity of the Green function is situated at
$(i=1,j=1)$.  We have introduced a cutoff in $r_1$ equal to
$\epsilon$.  We choose it to be a quarter of the lattice spacing,
i.e., $1/4$.  The reason for introducing the three other radii $r_2$,
$r_3$, and $r_4$ is that the Fourier transform makes the lattice
periodic.  The three additional radii signify the mirror image of the
singularity resulting from one reflection --- we do not introduce
further reflections since, with our choice of parameters, their effect
is negligible. The deformation field $u$ and the Green fuction, $G$
were then Fourier transformed, and Eq.\ (\ref{eq:green1}) solved in
Fourier space.  The resulting force field was then Fourier transformed
back to real space.  To within the approximations we have made, the
force field is proportional to the normal stress field $\sigma_N$.  We
show in Figure \ref{fig:realsurfstress} the normal stress field
corresponding to the full contact of the fracture in Figure
\ref{fig:realsurf}.  The strong small-scale variations in the normal
stress distribution are consistent with the observations reported in
\cite{Me98}.

In Figure \ref{fig:wavesurf}, we show the wavelet analysis of the normal
stress field obtained for the granite surface.  The least-squares fit gives
a slope of $0.30\pm0.04$ corresponding to a Hurst exponent of $-0.20\pm0.04$.
Thus, the relation $H_\sigma=H-1$ is supported.

In order to study systematically the relation between the Hurst exponent of
the deformation field and that of the normal stress field, we have
generated artificial self-affine surfaces, using the Fourier method
\citep{Sa98}.  This allows us to generate and subsequently average our
results over many surfaces for each Hurst exponent, in practice 100
surfaces.  After obtaining the stress fields for each surface, we
analysed spatial correlations of both the surfaces and the stress
fields with two techniques: the average wavelet spectrum of
one-dimensional traces obtained by cutting the surface along lines and
the two dimensional Fourier spectrum of the full
surface. Figure~\ref{fig:wavesurf} presents the average wavelet
spectra of the surfaces and their corresponding stress field for the
synthetic fracture surfaces with the same Hurst exponent as the one
that is observed for fracture surfaces: $H=0.80$. We treated each
surface and its corresponding stress field as consisting of 2048
one-dimensional traces, and as there were 100 surfaces, our averages
are over $100\times2048$ one-dimenensional traces. The scaling of the
synthetic surfaces is in good agreement with that of the measured
surface. Computed full contact stress fields of both types of surfaces
are also in good agreement supporting the relation: $H_\sigma=H-1$.
The two dimensional Fourier spectrum is computed from the two
dimensional Fourier transform of the surface and is expected to scale
for self-affine surfaces as \citep{Sa98}:
\begin{equation}
P(|{\bf k}|)\sim |{\bf k}|^{-2-2H}
\label{eq:2dfs}
\end{equation}

\begin{figure}
  \figurewidth{35pc}
  \figbox*{}{}{\includegraphics[width=25pc,angle=270]{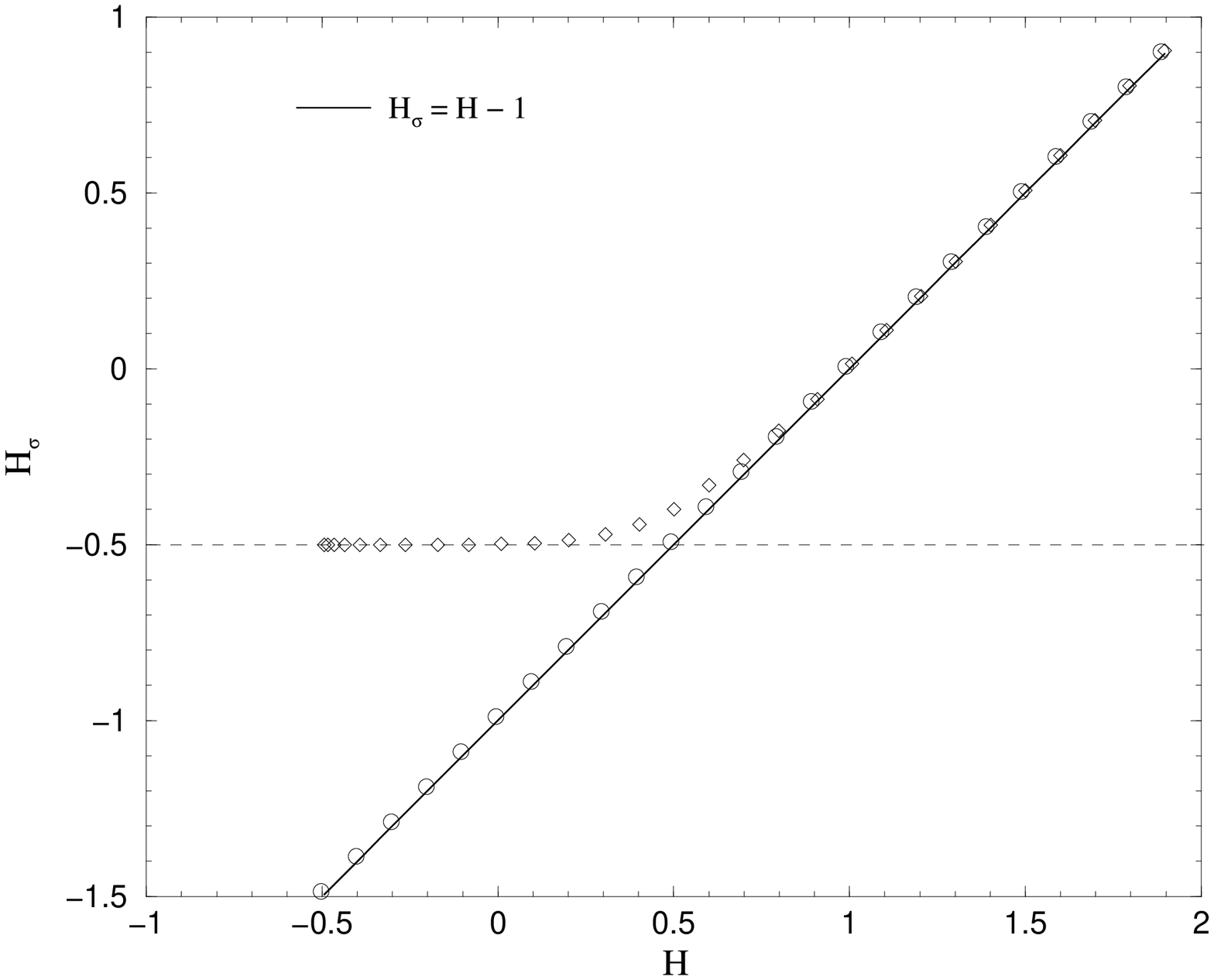}}
  \caption[]{\label{fig:hvshsigma} Hurst exponent of stress field,
  $H_\sigma$ as function of Hurst exponent of deformation field $H$.
  The data are based on averaging over several samples of artifially
  generated rough surfaces ($\diamond$ for 1D wavelet spectrum and
  $\circ$ for 2D Fourier spectrum).  The straight line is
  $H_\sigma=H-1$.  }
\end{figure}

 We generalize the analysis for different Hurst exponents that
 describe different spatial correlations between asperities.  In
 Figure
\ref{fig:hvshsigma}, we show $H_\sigma$ as a function of $H$ for the
artificially generated $2048\times 2048$ surfaces and analyses with
both techniques.  The straight line corresponds to
\begin{equation}
H_\sigma=H-1\;.
\label{eq:hsh}
\end{equation}
We see that the numerical results and Eq.\ (\ref{eq:hsh}) are in
excellent agreement for the two dimensional Fourier spectrum.  The
agreement is good with the one dimensional technique only for
sufficiently large roughness exponents of the rough surface. We
emphasize that measurements of low roughness exponent have to be done
with two dimensional techniques \citep{Ha-Sc-Ba}.

The Hurst exponent is directly related to the spatial correlations of
the surface. Eq.~\ref{eq:2dfs} shows that for: $H=-1$ surfaces have a
flat spectrum that is are white noise with no spatial correlations of
the asperities. When the Hurst exponent $H$ increases, relative
magnitudes of low frequency modes also increase. Asperities are
smoother and the surface roughness appears more and more correlated at
large scales. Equation~(\ref{eq:hsh}) shows that the stress field can
be calculated approximately as a simple derivative of the deformation
field. Also, fluctations of the stress field are significantly higher
than the deformation field.

We thank H.\ Nazareno of the ICCMP for the opportunity to perform this
work in an excellent and friendly environment. This work was partially
funded by the CNRS PICS contract $\#753$ and the Norwegian research
council, NFR.  We also thank NORDITA for its hospitality.



\end{article} 
\end{document}